# Local Dielectric Spectroscopy of Nanocomposite Materials Interfaces

*Massimiliano Labardi*[(1)*], *Daniele Prevosto*[(1)], *Kim Hung Nguyen*[(1,2)], *Simone Capaccioli*[(1,2)], *Mauro Lucchesi*[(1,2)] *and Pierangelo Rolla*[(1,2)]

[1] INFM-CNR LR polyLab, Largo Pontecorvo 3, 56127 Pisa, Italy

[2] Dipartimento di Fisica "Enrico Fermi", Università di Pisa, Largo Pontecorvo 3, 56127 Pisa, Italy

ABSTRACT.

Local dielectric spectroscopy is performed to study how relaxation dynamics of a poly-vinyl-acetate ultra-thin film is influenced by inorganic nano-inclusions of a layered silicate (montmorillonite). Dielectric loss spectra are measured by electrostatic force microscopy in the frequency-modulation mode in ambient air. Spectral changes in both shape and relaxation time are evidenced across the boundary between pure polymer and montmorillonite sheets. Dielectric loss imaging is also performed, evidencing spatial variations of dielectric properties near to nanostructures with nanometer scale resolution.

---

[*] Electronic mail: labardi@df.unipi.it

## I. Introduction

In nanocomposite materials, the high surface/volume ratio of nanostructures is exploited to obtain unique properties not available in bulk materials. For example, nanostructures dispersed in a host polymeric matrix can produce a large variation of bulk mechanical, optical, and electrical properties of a polymer as a consequence of the variation of polymer properties at the wide interface area with the nanostructures. Macroscopic characterization can be used to determine the overall behavior of the material, due to the concurrent effects of a large ensemble of nanostructures, although the effect of each of the single nanostructures remains not accessible. On the other hand, knowledge of the influence of a given type of nanostructure would help to design and engineer more easily new materials with the desired properties. Therefore, characterization techniques able to address single nanostructures are strongly demanded to this purpose. In particular, the electric response to externally applied fields, which is at the basis of many properties like e.g. local conductivity and dielectric behavior, is of strong interest for instance in micro-, nano-, opto-electronics, and solar energy conversion applications [1].

Among electrical characterization techniques applied in materials science, dielectric spectroscopy (DS) has revealed especially successful to disclose structural properties of polymeric materials, like e.g. their glass transition and structural relaxation dynamics [2]. In such technique the material of interest is placed between the plates of a capacitor, and the complex impedance is measured as a function of frequency. A possible approach to obtain highly local measurements of electric response is to use the tip of an atomic force microscope (AFM) as one of the electrodes [3]. The extremely small capacitance typical of such system makes detection of induced currents extremely challenging [4], therefore local impedance spectroscopy turns out to have only limited applicability.

Sensitivity of local dielectric measurements can be enhanced by exploiting the capability of AFM to detect extremely small forces and force gradients. By this way, very small capacitances can be measured by virtue of the dependence of electrostatic force on capacitance. Based on this concept, local probe DS [5] was implemented by means of an AFM-based method named frequency-modulated electrostatic force microscopy (FM-EFM) [6]. In essence, an excitation voltage $V(t) = V_{DC} + V_{AC} \cos(\omega t)$ is applied to the conductive tip of an AFM, and the electrostatic interaction force $F_{el}(t)$ with the sample is measured by frequency-modulated AFM methods [7]. More specifically, the AFM tip is vibrated in direction $z$, nearly orthogonally to the sample, at its own instantaneous resonant frequency $f_{res}$, whereas an external force field $F$ produces a resonance frequency shift

$$\frac{\Delta f}{f_{res}} = -\frac{1}{2k}\frac{dF}{dz} \qquad (1)$$

where $k$ is the spring constant of the AFM cantilever force sensor. When the tip is not concerned by appreciable atomic forces, i.e. with tip/sample distance larger than about 2 nm, only the electrostatic force gradient $dF_{el}/dz$ can affect the resonant frequency. Therefore, in such case $\Delta f$ is a direct measurement of the local $dF_{el}/dz$. The detailed description of the FM-EFM method can be found in the literature [6].

To operate the local DS mode, the modulation $\Delta f(t)$ induced by the AC voltage is measured in amplitude and phase by dual-phase lock-in technique. The EFM signal detected at frequency $2\omega$ is given by [8]:

$$F_{z,2\omega} = \frac{1}{4}V_{AC}^2 \cos(2\omega t)\frac{dC}{dz} \qquad (2)$$

where $C$ is the capacitance of the tip/sample system, depending on the complex dielectric function of the medium $\varepsilon(\omega) = \varepsilon'(\omega) - i\,\varepsilon''(\omega)$. Second harmonic detection is chosen since $F_{z,2\omega}$ is only influenced by capacitive interactions and not, for instance, by the contact potentials between tip and surface. In general, from Eq. (2) we can derive that the second harmonic component of $F_{z,2\omega}$ is complex, i.e. $F_{z,2\omega} = F_{z,2\omega}' + i\,F_{z,2\omega}''$. The presence of an imaginary part reflects the existence of a phase lag, in the following also called loss angle, $\delta_v$, of $F_{z,2\omega}(t)$ with respect to the excitation voltage, originating from the dielectric loss mechanism due to orientational polarization relaxation [5]. Local probe DS by EFM was recently demonstrated on homogeneous poly-vinyl-acetate (PVAc) films under vacuum [5].

In this work we demonstrate local DS in ambient air, and preliminarly apply it to address the influence of inorganic nano-inclusions on structural relaxation of polymers. PVAc ultra-thin films (35 nm thick) with a dispersion of a layered silicate (montmorillonite, MMT) were studied in ambient air with controlled humidity at a range of temperatures close to glass transition of PVAc (bulk $T_g$ = 310 K). Materials and methods used for this work are described in Sect. II. Our investigation indicates local changes of dynamical properties in the case of our nanocomposite sample, that is documented by comparison of loss tangent spectra obtained on the pure polymer and close to MMT inclusions. Spatial maps of dielectric properties have also been recorded that can be useful to investigate the spatial extension of such modifications. Experimental results, consisting in both dielectric spectra and maps, are presented in Sect. III. Discussion of experimental results is then given in Sect. IV.

## II.    Materials and methods

Ultra-thin films of a nanocomposite material, based on PVAc of molecular weight $M_w$ = 167,000 have been produced on gold substrates. PVAc was chosen as a host polymer to better compare with previous literature [5]. Gold layers of 30 nm thickness were obtained by thermal evaporation on a glass disk previously evaporated with a ~5 nm adhesion layer of chromium. Gold was electrically contacted to ground by silver paste. A PVAc-based nanocomposite has been obtained by dispersion of a layered silicate into the polymeric matrix. Organo-modified montmorillonite clay (MMT, $Na_xAl_{2-x}Mg_xSi_4O_{10}(OH)_2 \cdot n\ H_2O$), obtained by cation exchange of sodium with alkyl ammonium ions [9], was used to have a good dispersion down to single layers of MMT. Ultra-thin films were prepared by spin coating a solution of PVAc (2.3% w/w) and MMT (7 $10^{-4}$ % w/w) in toluene, at 5500 rpm for 60s, and annealed under vacuum at 303K for at least 12 hours to remove residual solvent. Films with thickness of about 35 nm were obtained, showing extended areas containing a few or even single MMT sheets, where sharp transitions to regions composed by pure polymer were present. Large aggregates of MMT were not observed. Such situation was pursued to allow the study of the transition region between the polymer and the inorganic nanostructure within. In order to uncover the metal electrode beneath, a scratch of the film was produced by a steel cutter prior to measurements. AFM profiling of the scratch was used to measure both film and gold layer thickness on our samples.

In this work, a modified Veeco Multimode AFM was used (Nanoscope IIIa with ADC5 extension). Our microscope is operated in controlled atmosphere by a home-made enclosure [10], and at controlled sample temperature by a heater/cooler stage driven by a thermal application controller (TAC). Relative humidity was always kept at less than 10% for all reported measurements. Frequency-modulation EFM is implemented through an electronics extender module (Quadrex), while signal access for external processing is obtained through a signal access module (SAMIII). For electric excitation and acquisition of dielectric spectra, an external dual-phase lock-in amplifier (SRS SR830DSP) controlled through GPIB interface by a home-made LabView routine is used. Doped silicon (Nanosensors PPP-FMR) AFM cantilevers were used, with spring constant $k \sim 3$ N/m, resonant frequency $f_{res} \sim 70$ kHz, and guaranteed tip radius of less than 10 nm. Oscillation amplitude was calibrated as customary by AFM TappingMode$^{TM}$ amplitude/distance curves, while actual sample temperature was calibrated by a thermocouple placed in contact with the upper surface of the sample.

The adopted experimental procedure was as follows. In order to ensure constant tip/sample distance during electrical measurements, the microscope was operated in LiftMode$^{TM}$. Such mode consists in a double-pass technique: during the first pass surface profiling is obtained with the usual tapping-mode operation, which also provides a tapping-mode phase image able to discriminate materials with different mechanical properties, while during the second pass the same scan trajectory is tracked by adding a constant lift height. Therefore, during the second pass there is no intimate contact between tip and sample, while their mutual distance is maintained at the same constant value until the next line is scanned. In our setup, during the first pass application of electrical potential to the tip is inhibited by an automatic switch, to reduce charge transfer effects. Some other technical details concerning the use of lift-mode for electrical measurements with FM-EFM are also illustrated in Ref. [11]. In particular, the determination of the actual tip/sample distance during the lift stage is especially relevant for quantitative evaluations. We typically worked at an average tip/sample distance of 15 nm with an oscillation amplitude of 4 to 6 nm.

To measure dielectric properties, the tip was placed on the desired spot of the sample, and its scanning motion was stopped by setting a scanning size of 0 nm. Afterwards, the spectra acquisition routine was executed, that provides a sequence of electrical excitation frequencies, and acquires the amplitude and phase of the second harmonic component of the $\Delta f$ modulation due to the applied sinusoidal potential. Time averaging is also performed in order to reduce noise. A number of loss tangent spectra, $\tan \delta_v(\omega)$, on the same spot can also be averaged to further reduce noise. The alternating tapping and lift-mode passes (although the tip is actually at rest in the lateral direction) provides an update of tip/sample distance, that could deviate with time from its set value due to thermal or mechanical drifts that may occur for too long measurement periods. As a drawback, the measurement time available in our system, that is, the duration of each lift-mode pass, could not be longer than 10s, due to settings constraints. This sets a lower limit for usable excitation frequency to ~ 1 Hz.

Each of our loss tangent spectra is referred to a calibration (baseline) measurement obtained on the bare metallic substrate, to take into account and subtract all measurement system responses that are not due to the behavior of the sample. To this purpose, the tip is positioned on the region of the scratch where the metal electrode is uncovered. Such calibration is repeated at each temperature and is performed at the same distance from the metal substrate than during the measurement on the dielectric film. In this way, we compare two situations with the same geometrical arrangement, the only difference being the presence or not of the dielectric layer. As an example, if a dielectric measurement is performed at a tip/sample distance of 5 nm from the polymer surface, and the film thickness is 35 nm, the calibration measurement is performed on the metal region at a distance of 40 nm.

With a different procedure, loss angle imaging, reflecting dielectric properties at a fixed frequency, can be performed. A value of modulation frequency is chosen, and the lock-in amplitude and phase outputs are acquired to form an image during the lift-mode scan. Due to settings constraints in our setup, the slowest point-to-point sampling rate achievable during imaging is $\Omega \sim 25$ Hz, so that imaging of signals demodulated from a carrier frequency lower than $\sim 3\Omega$ leads inevitably to some image resolution loss. Therefore, modulation frequencies within the range 60 – 130 Hz were typically chosen for optimal performance. We have recorded images at the border region between dielectric film and metal substrate, in order to obtain a visualization of the different dielectric behavior between the film and the substrate itself. For instance, presence of contrast between polymer and metal in the dielectric phase image means that a phase lag is present on the polymer compared to the metal, that is always characterized by in-phase response in the used low-frequency range. Operation in lift-mode for the acquisition of such maps leads to changing the tip/substrate distance during the scan, therefore modifying the system geometry, unlike in the case of dielectric spectra acquisition described above. Operation in LinearMode$^{TM}$ could be used to keep constant the tip/substrate distance. Linear-mode performs a straight trajectory during the second pass, at a given lift height above the average height of the previously tracked topographic profile (this is also called constant height mode, as opposed to constant gap mode [12]). However, operating linear-mode at small separation from the polymer surface is likely to cause undesired tip contacts with the surface, especially when topography is characterized by steps like in the case of the scratch produced on our sample. For this reason, operation in lift-mode was preferred in our case.

### III.    Experimental results

Fig. 1(a) shows the tapping-mode AFM topographic map of our PVAc-based nanocomposite film with MMT inclusions, as measured for a temperature higher than the glass transition temperature. The metal substrate exposed by scratching the film is visible on the right. We choose to show an area with high concentration of nanostructures, as well as with a number of irregularities, in order to better illustrate different possible contributions to image contrast. When MMT sheets are located close or on top of the polymer film upper surface, they are visible as corrugations in the topographic image. On the contrary, if sheets are completely embedded in the film, the corresponding topography appears as smooth as on the pure polymer regions. Simultaneous tapping-mode phase imaging (Fig. 1(b)) easily allows identification of MMT inclusions, even when they are completely buried in the polymer. In this image, brighter areas correspond to stiffer materials, either the metal substrate or the MMT sheets, compared to the softer polymer that appears darker. The substrate can be distinguished from MMT by comparison with the topographic map, because of its lower height and more pronounced roughness with respect to MMT.

Electric images recorded during the lift-mode second pass are shown in Fig. 1(c)-(e). By applying only a DC bias, the frequency shift image of Fig. 1(c) was recorded, corresponding to a measurement of electrostatic force gradient as from Eq. (1). This image demonstrates a lateral resolution of local electric measurements as high as 30 nm, as shown by the line profile analysis reported in Fig. 1(f), where a sharp transition in the electric signal is visible although the corresponding topography is clearly flat. The comparison of topography and electric signal profiles demonstrates that the observed contrast is not due to topography artefacts [12] but it is of genuine electrical origin, and is due to the presence of a buried MMT inclusion. By applying an AC bias at $f_{\text{mod}} = 130$ Hz and acquiring the phase of the demodulated second harmonic component of $\Delta f$, according to Eqs. (1) and (2), a loss angle image related to the local dielectric loss of the film is obtained as shown in Fig. 1(d) at $T = 320.6$ K. The loss angle $\delta_v$ measured by FM-EFM is defined as

$$\tan \delta_v = \frac{\dfrac{dF_{z,2\omega}^{"}}{dz}}{\dfrac{dF_{z,2\omega}^{'}}{dz}} \qquad (3)$$

that by Eq. (2) is equivalent to:

$$\tan \delta_v = \frac{\dfrac{d^2 C^{"}}{dz^2}}{\dfrac{d^2 C^{'}}{dz^2}} \qquad (4).$$

A generally accepted model for the tip/sample capacitance in presence of a dielectric layer of thickness $h$ and dielectric function $\varepsilon$, and valid over a large enough distance ($z$) range, is [13]:

$$C(z,\varepsilon,h) = 2\pi\varepsilon_0 R \ln\left(1 + \frac{R(1-\sin\theta_0)}{z + h/\varepsilon}\right) \qquad (5)$$

where $R$ is the tip apex radius, $\theta_0$ the conical tip angle, and $\varepsilon_0$ the permittivity of vacuum.

The darker appearance of the polymer film in Fig. 1(d) represents a negative phase lag due to dielectric relaxation at the excitation frequency $f_{\text{mod}}$ compared to the in-phase behavior of the metal electrode. Such contrast increases at $T = 329.2$ K (Fig. 1(e)), while it is completely lost at $T = 337.8$ K (not shown). By careful analysis of Fig. 1(e) one can notice that some of the MMT inclusions show up brighter than the surrounding polymer (i.e. they show reduced phase lag), while others do not.

This qualitative behavior can be confirmed by comparison of loss tangent spectra between pure polymer and MMT regions. Figs. 2(a)-(c) show such a comparison at three different temperatures. Shift of the dielectric loss peak when temperature is changed towards the glass transition is evident. Using Eq. (1) and (2) and modeling the tip-film-substrate capacitance as in Ref. [13], we can evaluate the expected signal in our experimental conditions. In Fig. 3 we compare one of our local DS spectra, scaled in magnitude for better comparison, with one calculated starting from dielectric spectra of bulk PVAc, measured by conventional broadband dielectric spectroscopy. We observe that the relaxation dynamics on the nanocomposite ultra-thin film is consistent with a smaller $T_g$ compared to the bulk. Faster structural dynamics in PVAc has been also reported in Ref. [5] showing lower $T_g$ for the surface of a much thicker film (~ 1 μm), as well as to Ref. [14] showing lowering of $T_g$ for thinner films from conventional DS measurements. Spectra acquired on the MMT layer appears shifted towards lower frequency as compared to the ones on pure PVAc (Fig. 2), indicating an increased $T_g$ due to the presence of MMT. Differences between spectra on the pure polymer and the MMT/polymer regions are also plotted in Fig. 2, for a better comparison.

## IV. Discussion

As already mentioned, a careful modeling of the measurement process carried out by FM-EFM is needed to achieve a general understanding of local dielectric relaxation measurements on the local scale, as well as to compare with bulk dielectric data. Contrarily to what typically reported in the literature [11,13] we do not measure distance dependency of electrostatic force gradient in order to infer the value of tip radius $R$ by fitting, since we observe from modeling results that the loss angle $\delta_v$ is not very influenced by $R$. Typical parameters used in our calculations with Eqs. (4) and (5), which reflect our experimental conditions, are $R = 10$ nm, $\theta_0 = \pi/6$, $h = 35$ nm, and $z_0 = 15$ nm. The dielectric function has been modeled by using a Havriliak-Negami equation:

$$\varepsilon_r(\omega) = \varepsilon_\infty + \frac{\Delta\varepsilon}{(1+(i\omega\tau)^{1-\alpha})^\beta} \qquad (6)$$

where $\varepsilon_\infty$ is the completely relaxed dielectric constant with respect to the structural relaxation process, $\Delta\varepsilon = \varepsilon_0 - \varepsilon_\infty$ is the dielectric strength, $\tau$ is the relaxation time, $\alpha$ and $\beta$ are the width and the asymmetry parameters of the dielectric structural peak [15].

In Figs. 2 (a) to (c), fitting curves of tan $\delta_v$ at different temperatures are reported as continuous lines.

The fitting has been carried out by minimizing the differences between experimental data and the fitting function obtained by combining Eqs. (4), (5) and (6). For the reported data analysis, $\varepsilon_\infty$ has been set to 3.6, which is the value in the bulk, and one or two other parameters (typically $\alpha$ and/or $\beta$) were fixed in order to obtain an overall reasonable reproduction of the experimental spectra by best fitting of the remaining ones. All fit parameters obtained are reported in Table I. In addition, the Kohlrausch-Williams-Watts parameter $\beta_{KWW}$, i.e. the exponent of the stretched exponential relaxation function used to represent data in the time domain, has been reported, by calculating its value from Havriliak-Negami parameters according to the relation $\beta_{KWW} = ((1-\alpha)\beta)^{0.813}$ [16].

The comparative analysis of spectra shows that the presence of MMT slightly slows down the structural relaxation of the polymer, and also modifies the structural peak shape. The slowing down can be observed in Fig. 4, where the frequency of the dielectric loss peak maximum, $f_{max}$, is reported as a function of the inverse of temperature [17]. We observe that the structural peak, as measured in regions containing both polymer and MMT, is more asymmetric than in the region with pure polymer. In presence of MMT, the $\alpha$ parameter of Eq. (6) decreases from 0.32 to 0.21 and the $\beta$ parameter decreases from 0.66 to 0.34, obtained in the case of the lowest temperature. Also the $\beta_{KWW}$ parameter, that matches the bulk values (0.52) at the same temperatures [18] when measured in correspondence of the polymer, is markedly reduced when MMTs are approached. This behaviour corresponds to a broadening of the structural peak, probably reflecting a more heterogeneous nature of the process in regions containing both MMT and polymer. The broader shape of the structural peak is accompanied by a relative increase of the slow modes of the process with respect to the fast ones, in agreement with the observation of the overall slowing down of structural dynamics, shown by the decrease of $f_{max}$. Aside from the dynamics slowing down in presence of MMT, there seems to be no evidence of activation energy modification by the MMT, that is indicated by the slope of Arrhenius plots (Fig. 4).

Inferences on the dynamics of polymer chains at the interface with inorganic nano-inclusion have been often performed starting from dielectric measurements on a macroscopic portion of the sample (for example Refs. [19,20,21,22,23]). In particular, in presence of a strong interaction between polymer and the nano-inclusion, slowing down of the interfacial dynamics is usually argued [20,21,22,23]. PVAc and MMT should be characterized by such a strong interaction, due to the polar nature of the acetate group, and as confirmed by the good adhesion of the two components deduced from topographic AFM images. Consequently, the results herein reported are, to the best of our knowledge, the first direct proof of the slowing down of polymer dynamics at the interface with an interacting MMT layer, measured with nanometric resolution.

Finally, we find that the measured dielectric strength is much smaller (about a factor 2.5 in our case) than the one of the bulk. The origin of such discrepancy can only partly be explained by confinement effects as reported in Ref. [14] on ultra-thin films by conventional DS, and is currently under investigation.

## V. Conclusions

Dielectric properties of a ultra-thin film formed by a PVAc/MMT nanocomposite material have been measured to investigate the modification of polymer mobility close to the inorganic inclusion. Local dielectric properties have been measured in ambient air as a function of position and temperature with an AFM operated in FM-EFM mode. Dielectric relaxation imaging on nanometric scale is performed at constant temperature demonstrating nanometer scale resolution of local DS on heterogeneous materials. Dielectric spectra acquired across the boundary between pure PVAc and MMT show a broadening of the structural peak and a slowing down of structural dynamics due to the presence of the nanostructures. Such results directly verify, with local scale measurements, the effect on polymer mobility due to well dispersed and interacting MMT layers, previously argued by macroscopic measurements on nanocomposite materials.

**Acknowledgements**

Financial support from the INFM-CNR "SEED 2007" project and the MIUR-FIRB project RBNE03R78E ("NANOPACK") is acknowledged. We thank Pasqualantonio Pingue (NEST-CNR and SNS) for sample deposition facilities, and Marco Bianucci for technical support. Image processing was performed by WSxM software [24].

**Figure and table captions.**

**Figure 1**: PVAc thin film with MMT inclusions. a) Topography (grey tones represent the $z$ range, 0 to 76 nm). b) Tapping mode phase. c) Frequency shift under 2V DC bias ($\Delta f$ range –162 to –179 Hz). d) Dielectric loss at $T$ = 320.6 K ($\delta_v$ range 4.7 deg). e) Dielectric loss at $T$ = 329.2 K ($\delta_v$ range 4.1 deg). f) Line profile of topography and frequency shift images at the position indicated by the horizontal stroke in (a) and (c). A constant frequency shift (+170 Hz) has been added for representation purposes.

**Figure 2**: Dielectric spectra recorded on PVAc (black squares), MMT (red rhombs), and their difference (green squares), at three different temperatures: (a) 319.9 K, (b) 322.8 K, (c) 327.5 K. The measurement region is shown in (d) by a tapping-mode phase image (scan size 2 x 1 $\mu m^2$).

**Figure 3**: Dielectric spectrum (tan $\delta_v$) of the used PVAc polymer in bulk (open symbols), compared to a local dielectric spectrum (closed symbols) of the 35 nm thick film at the same temperature (323 K). The latter data were multiplied by a factor 2.6 for better comparison.

**Figure 4**: Arrhenius plot of relaxation frequency for the pure polymer (open symbols) and polymer-MMT region (closed symbols).

**Table I.** Fit parameters at different temperatures of the structural peak for measurements on the pure polymer and polymer-MMT regions.

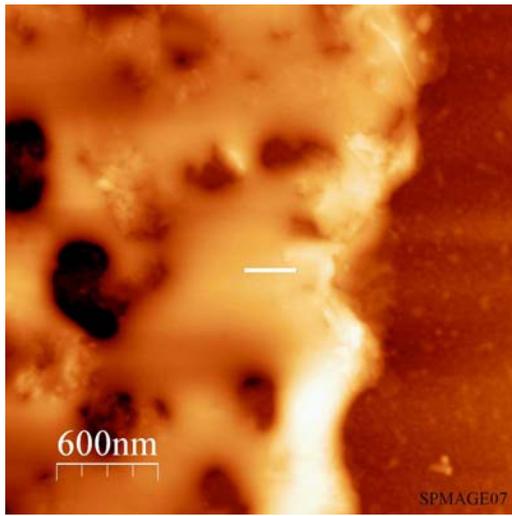 (a)
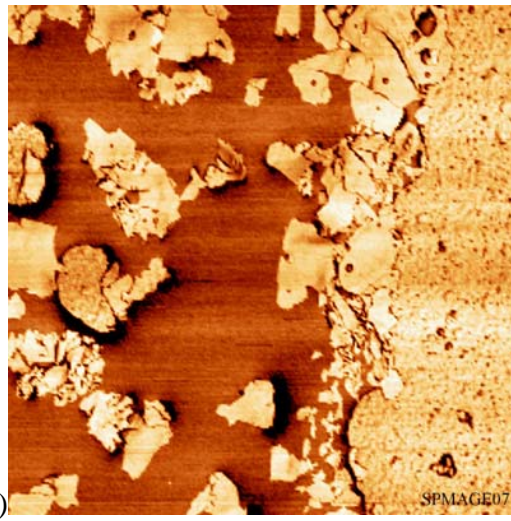 (b)
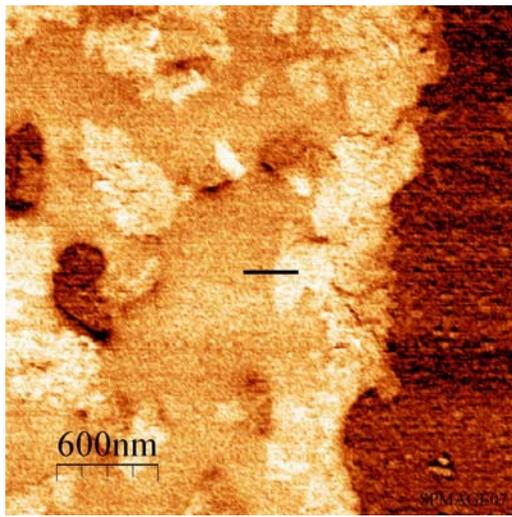 (c)
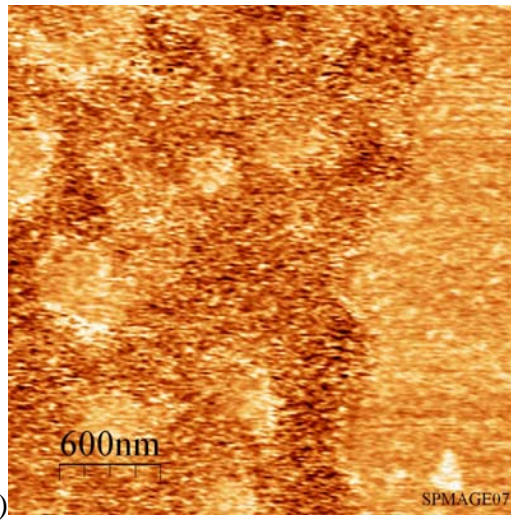 (d)
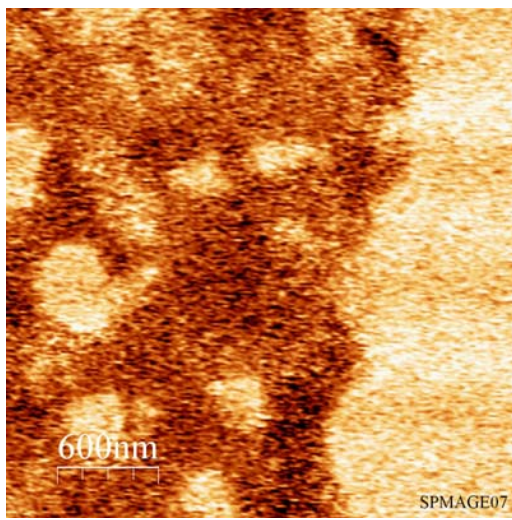 (e)
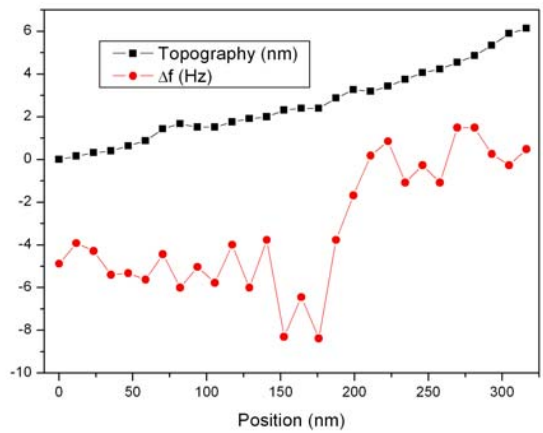 (f)

Figure 1.

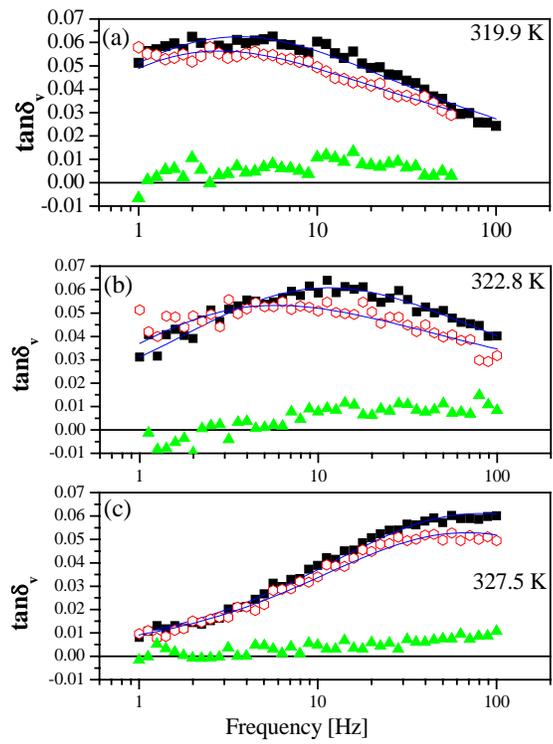
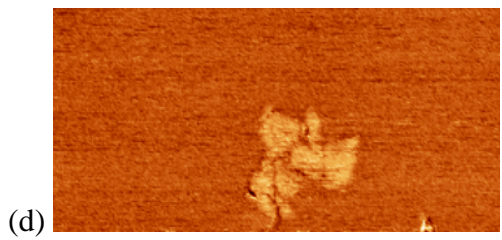

Figure 2.

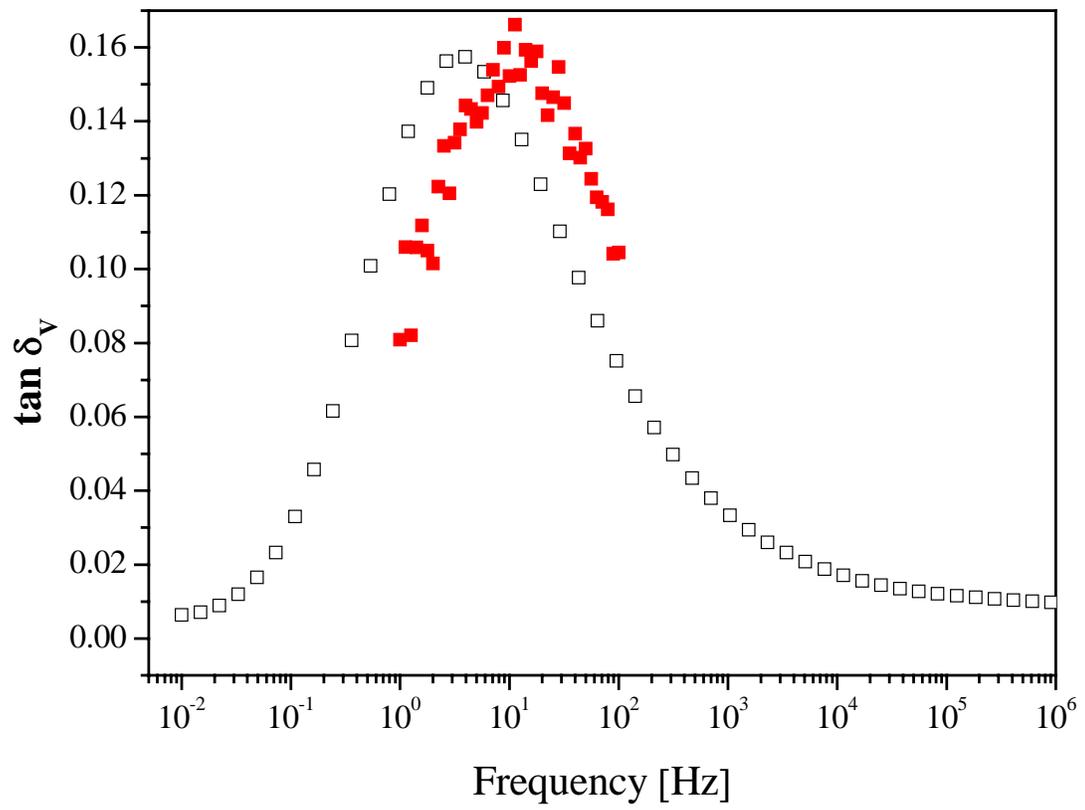

Figure 3.

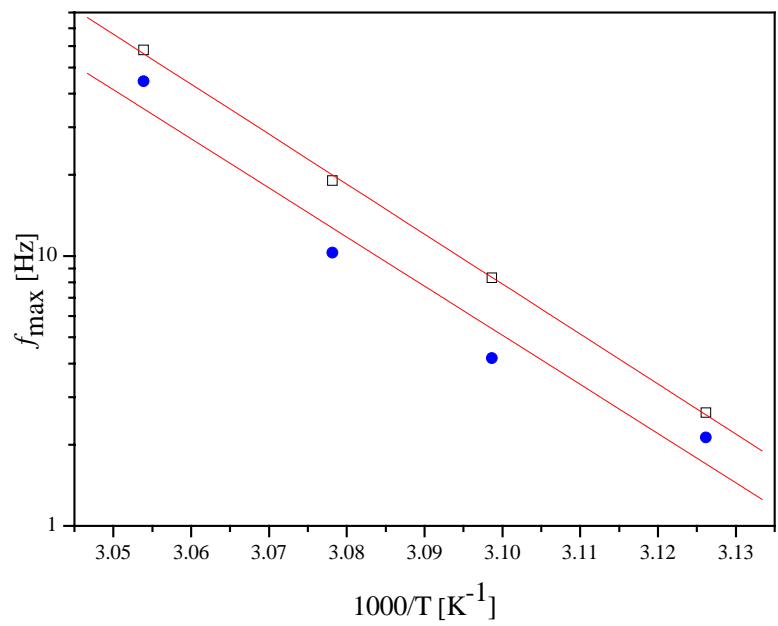

Figure 4.

| Material | $T$ [K] | $\Delta\varepsilon$ | $\varepsilon_\infty$ | $\alpha$ | $\beta$ | $\tau$ [s] | $\beta_{KWW}$ |
|---|---|---|---|---|---|---|---|
| PVAc | 319.9 | 1.01 | 3.6 | 0.32 | 0.66 | $1.05 \cdot 10^{-1}$ | 0.52 |
| | 322.8 | 0.98 | 3.6 | 0.32 | 0.66 | $3.31 \cdot 10^{-2}$ | 0.52 |
| | 324.9 | 1.00 | 3.6 | 0.32 | 0.66 | $1.44 \cdot 10^{-2}$ | 0.52 |
| | 327.5 | 0.98 | 3.6 | 0.32 | 0.66 | $4.73 \cdot 10^{-3}$ | 0.52 |
| PVAc + MMT | 319.9 | 0.94 | 3.6 | 0.21 | 0.45 | $1.81 \cdot 10^{-1}$ | 0.43 |
| | 322.8 | 1.03 | 3.6 | 0.21 | 0.36 | $1.17 \cdot 10^{-1}$ | 0.36 |
| | 324.9 | 1.01 | 3.6 | 0.32 | 0.26 | $9.35 \cdot 10^{-2}$ | 0.24 |
| | 327.5 | 1.01 | 3.6 | 0.20 | 0.35 | $1.15 \cdot 10^{-2}$ | 0.36 |

Table I.